\documentclass[article,a4,referee]{aa}

\usepackage{graphicx} 
  
\usepackage{txfonts} 
 
\begin{document} 
 
\title{A Lower-Limit Flux for the Extragalactic Background Light.} 
\author{Tanja M. Kneiske\inst{1} \and Herv\'e Dole\inst{2}} 
 
\institute{Institute f\"ur Experimentalphysik, University of Hamburg,  Luruper Chaussee 149, D-22761 Hamburg, Germany 
\and Institut d'Astrophysique Spatiale, Universit\'e Paris Sud 11 \& CNRS (UMR 8617), b\^at 121, F-91405 Orsay, France} 
\offprints{T.M. Kneiske \email{tanja.kneiske@desy.de}} 
 
\date{Received / Accepted }
 
\abstract
{The extragalactic background light (EBL) contains information about
  the evolution of galaxies from very early times up to the present.
  The spectral energy distribution is not known accurately, especially
  in the near- and mid-infrared range.  Upper limits and absolute
  measurements come from direct observations which might polluted by
  foreground emission, while indirect upper limits can also be set by
  observations of high energy gamma-ray sources. Galaxy number counts
  integrations of observable galaxies, missing possible faint sources,
  give strict lower limits.}
{A model is constructed, which reproduces the EBL lower limit flux.
  This model can be used for a guaranteed, minimum correction of
  observed spectra of extragalactic gamma-ray sources for
  extragalactic absorption. }
{A forward evolution model for the metagalactic radiation field is
  used to fit recent observations of Spitzer, ISO, Hubble and
  GALEX. The model is applied to calculate the Fazio-Stecker relation,
  and to compute the absorption factor at different redshifts and
  corrected blazar spectra.}
{A strict lower-limit flux for the evolving extragalactic background
  light (and in particular the cosmic infrared background) has been
  calculated up to redshift of 5. The computed flux is below the
  existing upper limits from direct observations, and in agreement
  with all existing limits derived from very-high energy gamma-ray
  observations. The corrected spectra are still in agreement with
  simple theoretical predictions. The
  derived strict lower-limit EBL flux is very close to the upper
  limits from gamma-ray observations. This is true for
  the present day EBL but also for the diffuse flux at higher redshift.}
{If future detections of high redshift gamma-ray sources require a
  lower EBL flux than derived here, the physics assumptions used to
  derive the upper limits have to be revised.  The lower-limit EBL
  model is not only needed for absorption features in AGN and
  other gamma-ray sources but 
  is also essential when alternative particle processes are
  tested, which could prevent the high energy gamma-rays from being
  absorbed. It can also be used for a quaranteed interaction of cosmic-ray
  particles. The model is available online.}
  

\maketitle

\section{Introduction} 
Diffuse extragalactic background radiation has been observed over a
broad range of the energy spectrum from radio to high energy
gamma-rays. A main contribution at almost all wavelength (except for
the Cosmic Microwave Background) are faint point sources (sometimes
unresolved), emitting in the energy band of interest.  Therefore, the
extragalactic background radiation turns out to be a good tool to
study global parameters of source populations and universal
physics. The optical to infrared extragalactic diffuse radiation, also
called extragalactic background light (EBL), is the relic emission of
galaxy formation and evolution, and is produced by direct star light
(UV and visible ranges) and light reprocessed by the interstellar dust
(infrared to sub-millimeter ranges). Minor contribution may include
genuine diffuse emission (e.g. galaxy clusters, Chelouche, Koester\&
Bowen 2007) or other faint sources (e.g. Population III stars, Raue,
Kneiske \& Mazin 2008).
 
While it is possible to measure extragalactic diffuse emission in the
sub-mm range, (Puget et al., 1996; Hauser et al., 1998; Hauser \& Dwek
2001), the EBL is difficult to measure directly in the infrared
because of strong foreground contamination. Thus, upper limits have
been derived by observing the isotropic emission component (see Hauser
\& Dwek, 2001 and Kashlinsky 2005 for reviews, as well as Lagache et
al., 2005 and Dole et al., 2006). Lower limits can be derived by using
integrated galaxy number counts which has been improved during the
last years by sensitive telescopes like {\it Spitzer} (Werner et al.,
2004), to get lower limits; this technique gives good constraints at
wavelengths shorter than 24 microns. At larger wavelengths, higher
confusion and lower sensitivities lead to very small lower limits. To
overcome the poor constraints at far-infrared wavelengths, a stacking
analysis of near- and mid-infrared sources is used (e.g. Dole et al.,
2006), to significantly resolve the cosmic infrared background (CIB),
leading to constraining lower limits.

Other constraints on the EBL are coming from the study of distant
sources of very high energy gamma-ray emission.  High-energy gamma
rays traveling through intergalactic space can produce
electron-positron pairs in collisions with low energy photons from the
extragalactic background light (Nikishov 1962, Goldreich \& Morrison
1964, Gould \& Schreder 1966, Jelley 1966).  Despite this effect,
Cherenkov telescopes have been discovered a great number of
extragalactic high energy gamma-ray sources at unexpected large
redshift. The discovery of 3C279 by the MAGIC telescope collaboration
(Albert et al., 2008) shows that 80-500 GeV gamma-rays photons can
travel distances from redshift z=0.536 without being too heavily
absorbed.  From the observation of H2356 - 309 and 1ES 1101-232, the
H.E.S.S. collaboration derived an upper limit for the EBL between 1
and 4 micron (Aharonian et al. 2006) , which is very close to the
optical number counts by the {\it Hubble Space Telescope} (Madau \&
Pozzetti, 2000). They verified their result in Aharonian et
al. (2007a) with the BL Lac 1ES 0347-121 and extended their limit to
the mid infrared using 1ES 0229+200 Aharonian et al. (2007b).  The
caveat is that the upper limit strongly depends on the assumption of
the intrinsic blazar spectrum.


Different types of models for the EBL flux have been developed.  The
simplest method (backwards evolution) extrapolates present day data or
template spectra to high redshift in a certain wavelength range (for
the most recent ones: Chary \& Elbaz 2001; Malkan \& Stecker 2001;
Totani \& Takeuchi 2002; Lagache et al., 2003, 2004; Xu et al., 2003;
King et al., 2003, Stecker et al., 2006; Franceschini et al.,
2008). Cosmic chemical evolution models self-consistently describe the
temporal history of globally averaged properties of the Universe (Pei,
Fall \& Hauser 1999) but falls short when it comes to comparisons with
data of individual galaxies.  Semi-analytical models are invoking
specific hierarchical structure formation scenarios to predict the
Metagalactic Radiation Field (MRF, i.e. the EBL at various redshifts)
(e.g. Balland et al., 2003; Primack 2005). The model used in this
paper is an updated version of the Kneiske et al. (2002, 2004) forward
evolution model. Simple stellar population models are used to describe
the evolution of stars in the universe from their very first formation
up to the presents.  Not only the physics of stars but also the
composition and spatial distribution of the interstellar medium are
taken into account.

\begin{figure}[t] 
\includegraphics[width=0.48\textwidth]{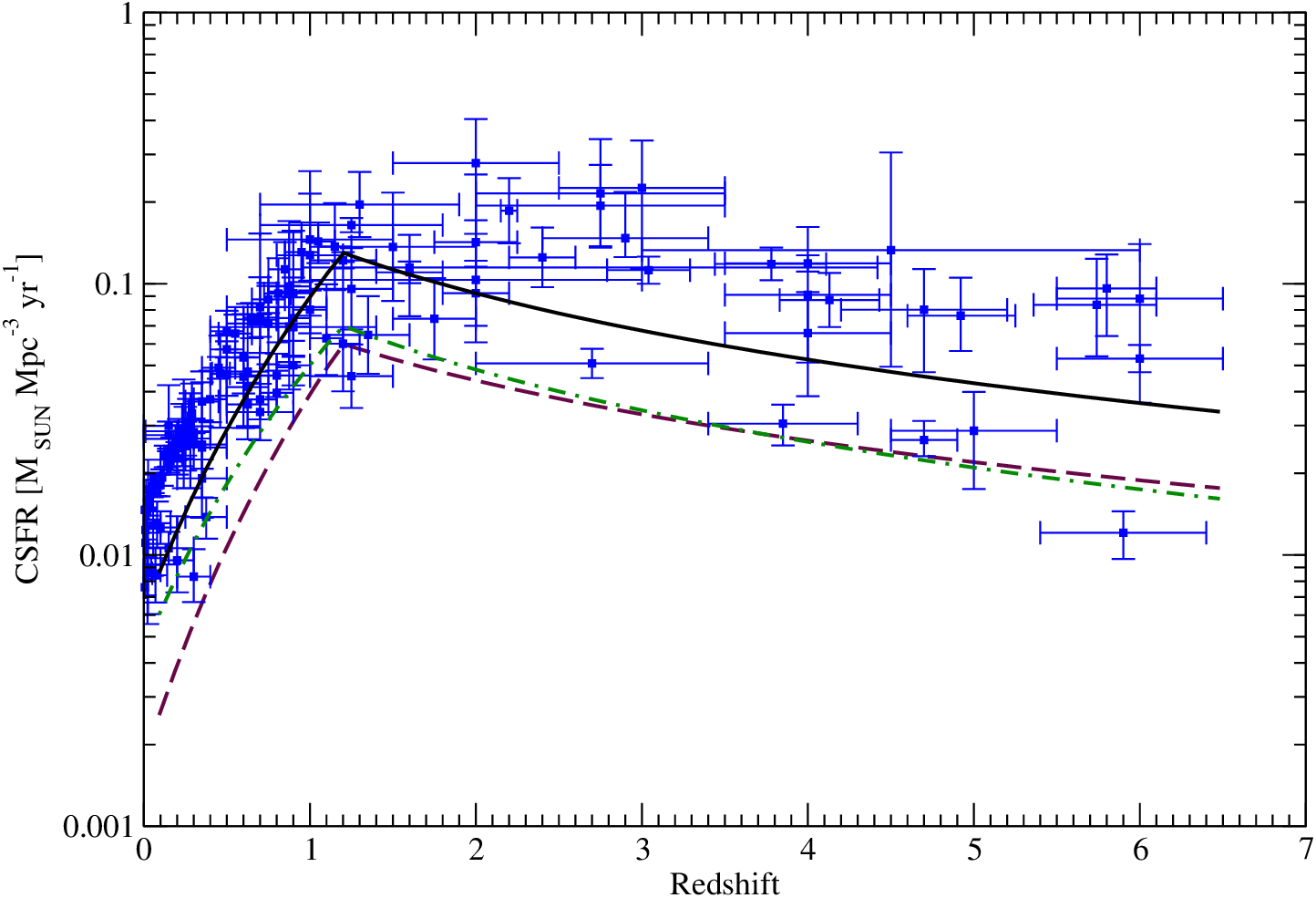} 
\caption{Comoving cosmic star formation rate. The data are taken from
  \cite{hopkins2006}.  The solid line shows the model total star
  formation rate, while the dashed and dashed-dotted line accounts the
  contribution from dust-poor and dust-rich regions respectively.}
\label{abb:SFR} 
\end{figure}

In this work lower-limit EBL data are used, to derive a lower-limit
EBL flux model.  In the next section, the data and their uncertainties
are discussed. The minimum EBL flux model is derived in the third
section by choosing parameters for the global star formation and the
interstellar medium. The results are presented in the fourth section,
together with the resulting optical depth for gamma-rays in the
universe.  Throughout this paper, a cosmology with $h=0.72$,
$\Omega_{\rm M} = 0.3$ and $\Omega_{\Lambda} = 0.7$ is adopted.

\begin{figure}[t] 
\includegraphics[width=0.45\textwidth]{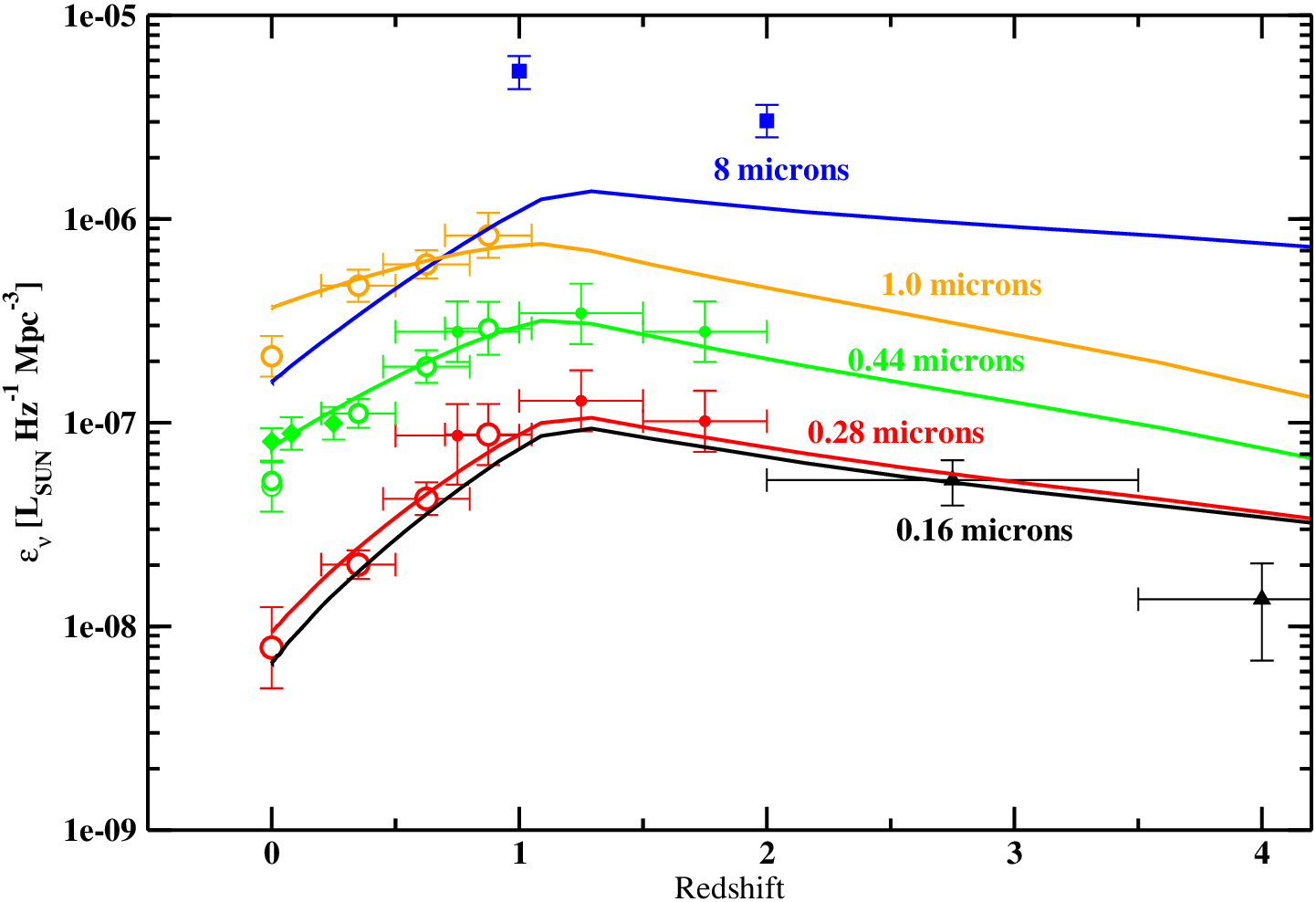} 
\caption{Comoving emissivity as a function of redshift. The lines are 
  calculated for the wavelength indicated in the figure and have to be 
  compared with the data points of the same color. Data come from: 
  \cite{ellis96,lilly96,connolly97,pozzetti98,caputi2007}.} 
\label{abb:emis} 
\end{figure}

\section{Current lower limits on the Cosmic Optical and Infrared Backgrounds} 
\label{sect:constraints_cib} 
 
Lower limits on the Extragalactic Background Light measurements are
reviewed briefly.  Most of them are derived from the integration of
number counts, not from direct measurements of surface brightness,
which are subject to strong foreground emission contamination.  This
method is based on the simple counting of detected galaxies on a given
sky area of a deep survey, a completeness correction, and the flux
integration of the number counts. Variance due to large-scale
structure may affect the results, and are usually taken into account
in the error bars. However, another source of uncertainty at
near-infrared wavelengths is the usually poor detected galaxy
statistics at large flux densities, and the subtraction of stars;
these uncertainties affect the number counts at high flux densities,
and can give different results when integrating them to get the
background lower limit.  Any model of the EBL should thus lie above
these observed limits. In the past not all EBL models meet this
criterion and are therefore not realistic and in contradiction with
the data.  The lower limit data are shown in Fig.~3 as data points
with the errors discussed below.

\subsection{Ultraviolet and visible EBL}
Counts and integration was done by \cite{xu2005} ({\it GALEX}); 
\cite{brown2000} and \cite{gardner2000} ({\it HST/STIS}); 
\cite{madau2000} and \cite{totani2001} ({\it HST/WFPC2}). 

\subsection{Near- and mid-infrared EBL} 
The integration of number counts on deep surveys done with the HST was
done by \cite{madau2000} and \cite{thompson2003,thompson2007} ({\it
  NICMOS}), and \cite{totani2001} ({\it SUBARU}).  

\cite{fazio2004a} obtained number counts with {\it Spitzer/IRAC} at
3.6, 4.5, 5.8 and 8.0~$\mu$m, and derived lower limits. These counts
have been confirmed by \cite{magdis2008} at these 4 wavelengths, and
by \cite{franceschini2006} at 3.6~$\mu$m.  At 8.0~$\mu$m, however,
\cite{franceschini2008} recomputed the counts at larger flux densities
with better statistics and re-integrated the whole number counts; they
claim that their integration gives a 50\% smaller value that
\cite{fazio2004a}; The value published by \cite{franceschini2008} will
be used as a lower value at 8.0~$\mu$m . In the same spirit, the
5.8~$\mu$m estimate would need to be recomputed.  At 3.6~$\mu$m,
\cite{levenson2008} integrated the extrapolated number counts (using
constraints from the image noise) and are getting close to the DIRBE
minus 2MASS value, giving an estimate of the CIB at this
wavelength. As a strict 3.6~$\mu$m lower limit, the \cite{fazio2004a}
value is used.  It should be noticed, however, that IRAC counts at
this wavelength may not be that reliable when integrated to give CIB
lower limits, despite the fact that number counts are very accurately
measured in deep surveys at faint flux densities (e.g agreement
between \cite{fazio2004a}, \cite{franceschini2006} and
\cite{magdis2008} at 3.6~$\mu$m).  Counts are contaminated by the
presence of bright and faint stars, as well as extended local
galaxies, biasing the measure at large and intermediate flux
densities, where deep surveys have very poor statistics; Large and
shallow surveys have better statistics, but the star contribution
subtraction might be inaccurate and can dominate the systematics
uncertainty. Nevertheless the data point will be included in our
analysis, where the error bars represent the large uncertainties.

In the mid-infrared, the counts by \cite{elbaz2002} 15~$\mu$m
using {\it ISOCAM} are used. At 24~$\mu$m with {\it Spitzer/MIPS}, the
counts by \cite{papovich2004}, \cite{marleau2004} and
\cite{chary2004}, \cite{rodighiero2006} are used.  At these wavelengths,
  there is not anymore a problem of star contribution (star spectra
  are far in the Rayleigh-Jeans regime) nor extended galaxy problem
  (point spread functions are larger than 6 arcseconds): the lower
  limits are reliable.
 
\subsection{Far-infrared and sub-millimeter EBL} 
Above 30~$\mu$m wavelength, another method than 
integrating the number counts is used, because individual detected 
far-infrared sources do not contribute more than 25\% to the 
background (e.g. \cite{dole2004,frayer2006}), except in the GOODS 
70~$\mu$m survey (about 60\% \cite{frayer2006b}). This method consists 
in stacking a longer-wavelength signal at the position of known short 
wavelength sources, and measure the resulted total flux, which is also 
a lower limit. At 70 and 160~$\mu$m, the lower limits of \cite{dole2006} 
obtained with a stacking analysis of {\it Spitzer/MIPS} 24~$\mu$m 
sources is used.  
The submillimeter {\it COBE/FIRAS} spectrum of direct 
detection comes from \cite{lagache2000}.

\begin{figure}[t] 
\includegraphics[width=0.49\textwidth]{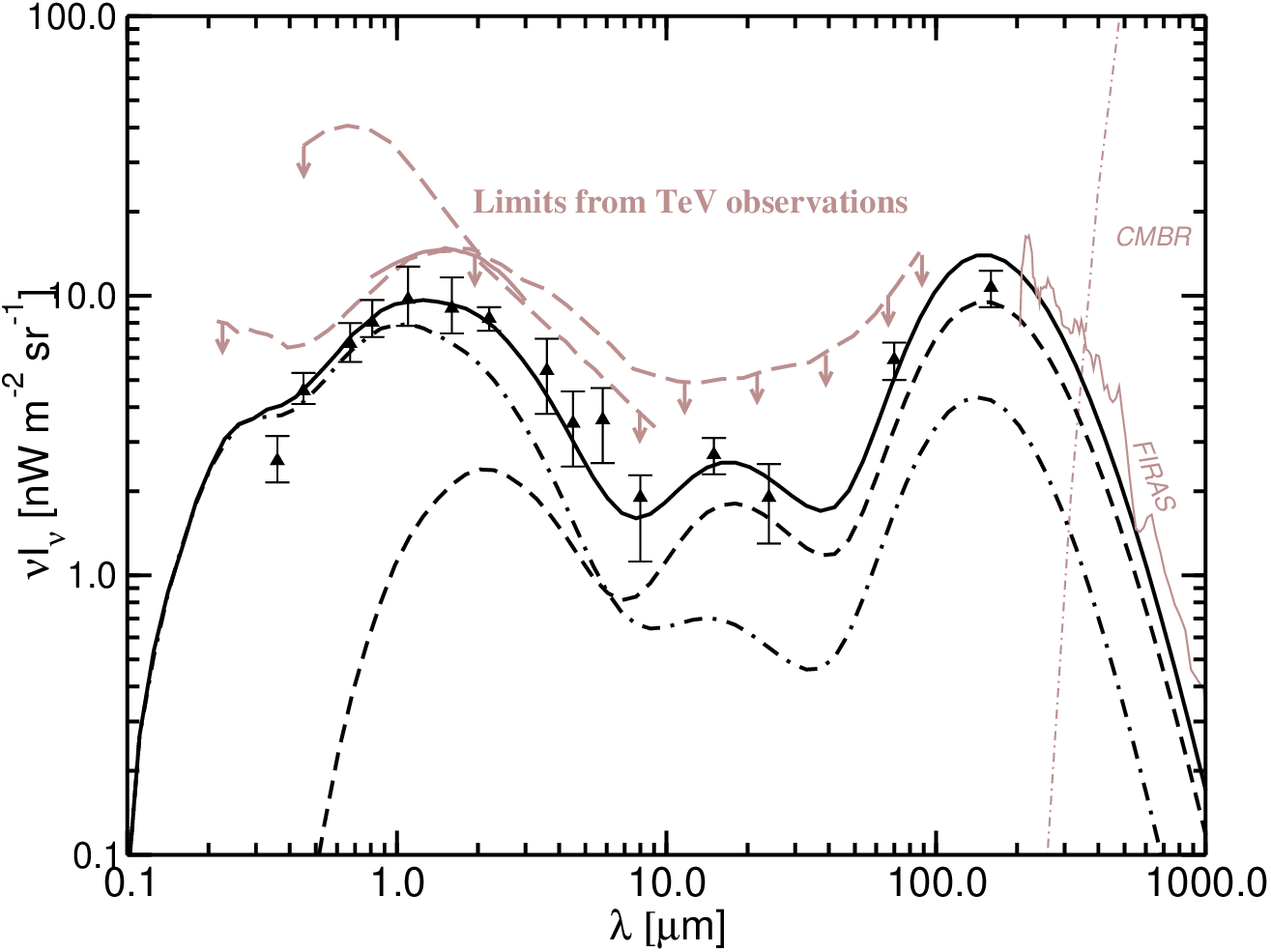} 
\caption{Extragalactic Background Light Spectral Energy Distribution.
  Data are lower limits (filled triangles), discussed in
  Sect.~\ref{sect:constraints_cib}.  The total model flux is shown as
  black solid line, together with the contribution from dust rich
  (dashed line) and dust poor star forming regions (dot-dashed line).
  The red dashed line are model-dependant upper limits on the EBL as
  derived from high energy blazar observations (Aharonian et
  al. (2006), Aharonian et al. (2007), Albert et al. (2008) and Mazin
  \& Raue (2007). Other long-wavelength detections are plotted: the
  submillimeter EBL and the CMB.  }
\label{abb:EBL_LL} 
\end{figure}

\begin{table}[!b] 
\centering 
\caption{Model Input Parameters (Definitions see Kneiske et al. 2004)} 
  
 \begin{tabular}{lllll} 
 \hline \hline\ 
 & $\alpha$ & $\beta $ & $z_p$ & $ \dot{\rho_\ast}(z_p)$  \\ 
 &&&&[$M_\odot$ Mpc$^{-3}$ yr$^{-1}$]        \\ 
 \hline  
 
 &&&&\\ 
 \bf{Strict lower-limit model}  &&&& \\ 
 $SFR_ 
 {OPT}$   & 3.5 & -1.2 & 1.2 & 0.07 \\ 
 $SFR_{LIG}$   & 4.5 & -1.0    & 1.2 & 0.06 \\ 
 $f_{esc}=0$ &	   &	  &     &     \\ 
 $c_2 = 10^{-24}$ &&&&\\ 
      \hline 
  
 \end{tabular} 
 \label{tab1} 
 \end{table}

\section{Lower limit EBL model} 
In this section an EBL model is constructed which reproduces the EBL
flux lower limits from source counts.  The EBL model is described in
details in Kneiske et al. (2002) and the main features are summarized
below.  The idea is to describe cosmological stellar evolution using a
simple stellar population model depending on different stellar
masses. The cosmological evolution is set by an input comoving star
formation rate density (SFR). The model computes emissivities and the
EBL flux, which can be directly compared with observations at individual
wavelengths. Two different star forming regions are distinguished
phenomenologically: "optical" star forming regions with low extinction
due to the presence of dust ($E(B-V)=0.06$), and "infrared" star
forming regions with higher extinction aiming at reproducing the
emission properties of Luminous and ultra-luminous infrared galaxies
(LIRG and ULIRG; $E(B-V)=0.8$). For these two populations, spectral
energy distributions (SED) are generated using a spectral synthesis
model, adding a consistent model accounting for dust absorption and
reemission. Three components of dust are taken into account by
modified black body spectra with different temperatures.  The goal is
thus to fit the EBL observed lower limit, by adjusting the input SFR
and dust parameters.

\begin{figure*}[!ht] 
\includegraphics[width=0.90\textwidth]{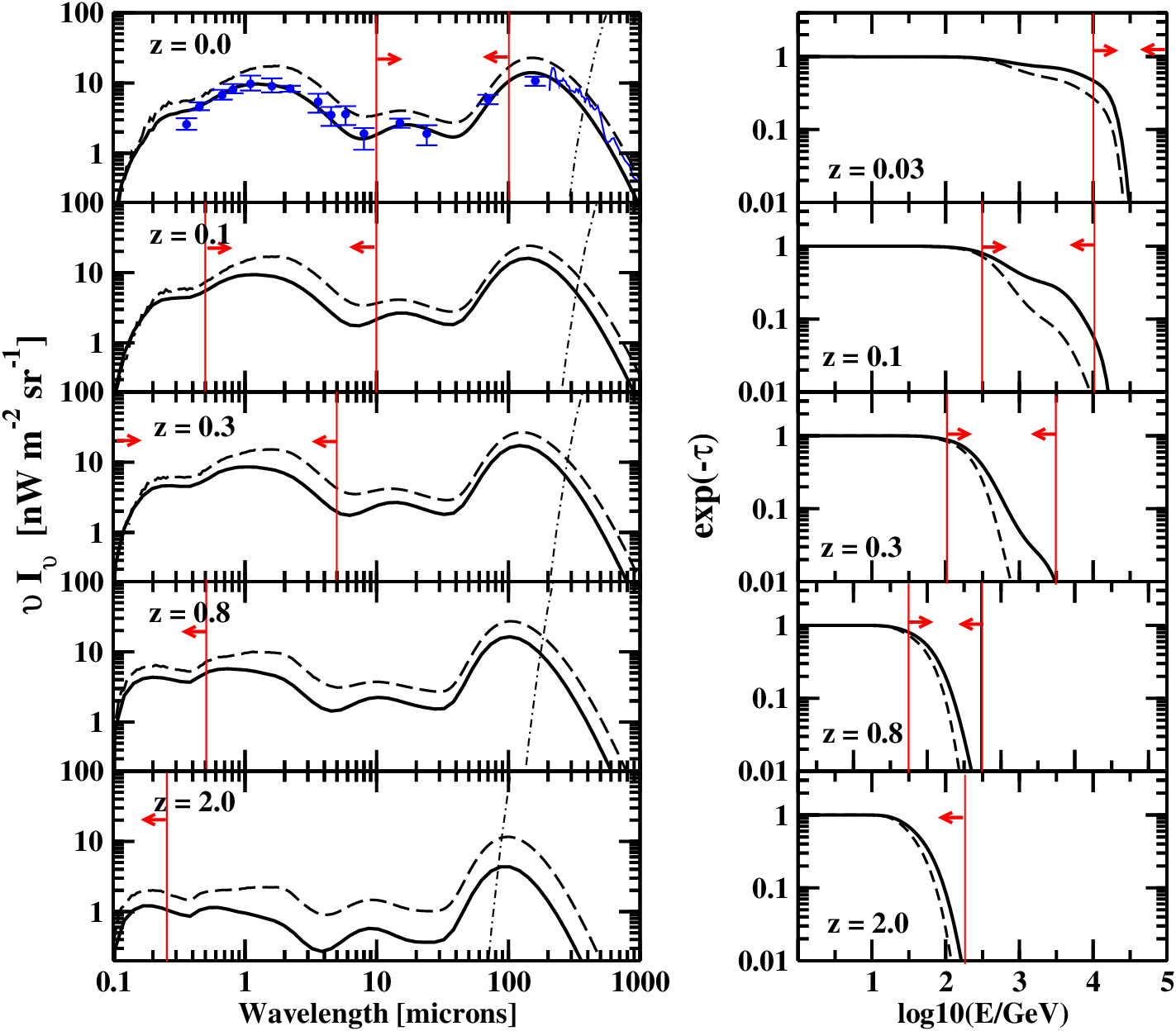} 
\caption{\emph{left:} Comoving flux of the extragalactic background light at five different redshifts.
  The solid line represents the lower-limit EBL introduced here, while
  the dashed line is the old ''best-fit'' model described in
  \cite{lit:kneiske2}. The spectral EBL region responsible for
    the cut-off at high energy is represented by thin vertical lines
    and arrows. \emph{right:} Extinction factor of gamma-rays as a function 
    of gamma-ray photon energy at five different redshifts.}
\label{abb:expTau} 
\end{figure*}

The EBL model flux has been fitted to the observed lower limits summarized in the
last section, by integrating the emissivities on the redshift range
zero to two. This takes into account the fact  that data are only able to
resolve galaxies up to a certain redshift, which depends on the flux
limit of the instrument and the survey. It is not possible to give the
exact maximum redshift for each survey, since the redshift is not
known automatically for each detected source. The chosen maximum
redshift of 2 seems a good average for most surveys taken into
account.  Our result is only weakly dependent on this parameter. The
model parameters have been chosen to minimize the $\chi^2$ between EBL
observed limits and the model. 

 

\section{Results and Discussion} 
 
\subsection{Cosmic Star Formation Rate and Emissivity} 
The model output cosmic star formation rate is shown in
Fig.~\ref{abb:SFR}; It is lower by a factor of 2 to 3 than the data,
compiled by \cite{hopkins2006}. This is not surprising, given the fact
that a lower limit EBL is used, which by definition is missing some
amount of emission. The shape, however, is consistent with the data.
 
Since the star formation rate is a model-dependent value which shows a
wide range of scatter, it is useful to compare the model emissivities
at different redshifts with integrated luminosity functions at various
wavelengths. As shown in Fig.~\ref{abb:emis}, the agreement between
optical ($\lambda \le 1~\mu$m) data and the model emissivity is good
for redshifts below 3.  
The model, however, is underestimating the emissivity at 8~$\mu$m by
a factor 3 to 5. The origin of this discrepancy might be twofold: 1)
the simplistic galaxies' spectral energy distribution used, lacking
detailed aromatic bands and very small grains continuum description;
and 2) a slight overestimation of the observed 8~$\mu$m emissivity,
obtained trough the rest-frame 8~$\mu$m luminosity function
integration (Caputi et al., 207) and an extrapolation to the infrared
bolometric luminosity density. Despite the care taken, this last
operation might slightly overestimate the emissivity. This might be
the reason why the model is not in strong disagreement with the EBL
shape at 8~$\mu$m (figure~3), despite a disagreement with the 8~$\mu$m
emissivity.


\subsection{Extragalactic background light (EBL)} 
The observed EBL lower limits (Sect.~\ref{sect:constraints_cib}) are
plotted in Fig.~\ref{abb:EBL_LL}, together with the model.  The model
reproduces the data well, keeping in mind that a physical model has
been used instead of a functional fit, and that the minimum
$\chi^2$ has been used. Almost all EBL flux (wavelengths $0.3 \le
\lambda \le 160~\mu$m) comes from galaxies up to a redshift of two, as
expected (e.g. Lagache, Puget, Dole, 2005). There is no significant
change in the computed EBL spectrum when including emission from
redshifts above 2, since the cosmic star formation rate drops by half
an order of magnitude. The robustness of our EBL
derivation is checked by integrating the emissivities up to a redshift of $z=5$:
this doesn't change the final result by more than 4\%.  The optical
and infrared EBL are dominated by their respective components (optical
and infrared galaxies), and the transition region between both
contributions, located around 5 microns, can be probed by {\it
  Spitzer}. The 5.8 micron data point lies above our model flux by
more than 1 $\sigma$. As discussed in
Sect.~\ref{sect:constraints_cib}, this point might suffer from a poor
statistics. At 8 micron, the new estimate of Franceschini et
al. (2008) lies on our model, but Fazio et al. (2004) estimate is
higher. While a consistent new estimate of all IRAC points would be
needed, it is possible yet to conclude if this discrepancy is a
common feature of EBL models (see also Franceschini 2008; Primack,
Gilmore \& Somerville 2008), and/or if the data points around 5
microns are overestimated (this last possibility cannot be ruled
  out, as discussed in Sec.~\ref{sect:constraints_cib}).  Finally,
our EBL model lies below the observed upper limits derived from
gamma-ray observations, as expected.

\subsection{EBL and $\gamma$-ray absorption at high redshift}

The lower limit EBL model can be used to calculate the optical depth
for photon-photon pair production. The effect is largely important in
extragalactic sources like blazars (Salamon \& Stecker 1998, Primack
et al. 1999, Kneiske et al. 2004) or gamma-ray bursts. The absorption
can result in a drastic change of the high energy spectrum or even
make it impossible to observe the source at all at gamma-ray energies.
The effect of absorption for extragalactic gamma-ray sources at
different redshift is shown in Figure~4. The EBL flux is plotted next
to the absorption factor $exp(-\tau)$ at the same redshift.  The
spectral region of the EBL flux responsible for the so cut-off region
is indicated by vertical red lines and arrows. The cosmic microwave
background is also plotted as dot-dashed line on the right of the EBL
flux diagram.  The results of our new lower-limit EBL model are
compared with the so called ''best-fit'' EBL model from
\cite{lit:kneiske2}.  It is clearly visible that a lower EBL flux is
leading to an absorption closer to one, which means lesser absorption
of gamma-ray photons in the cut-off region.

\subsection{Fazio-Stecker Relation}

\begin{figure*}[t] 
\includegraphics[width=0.75\textwidth]{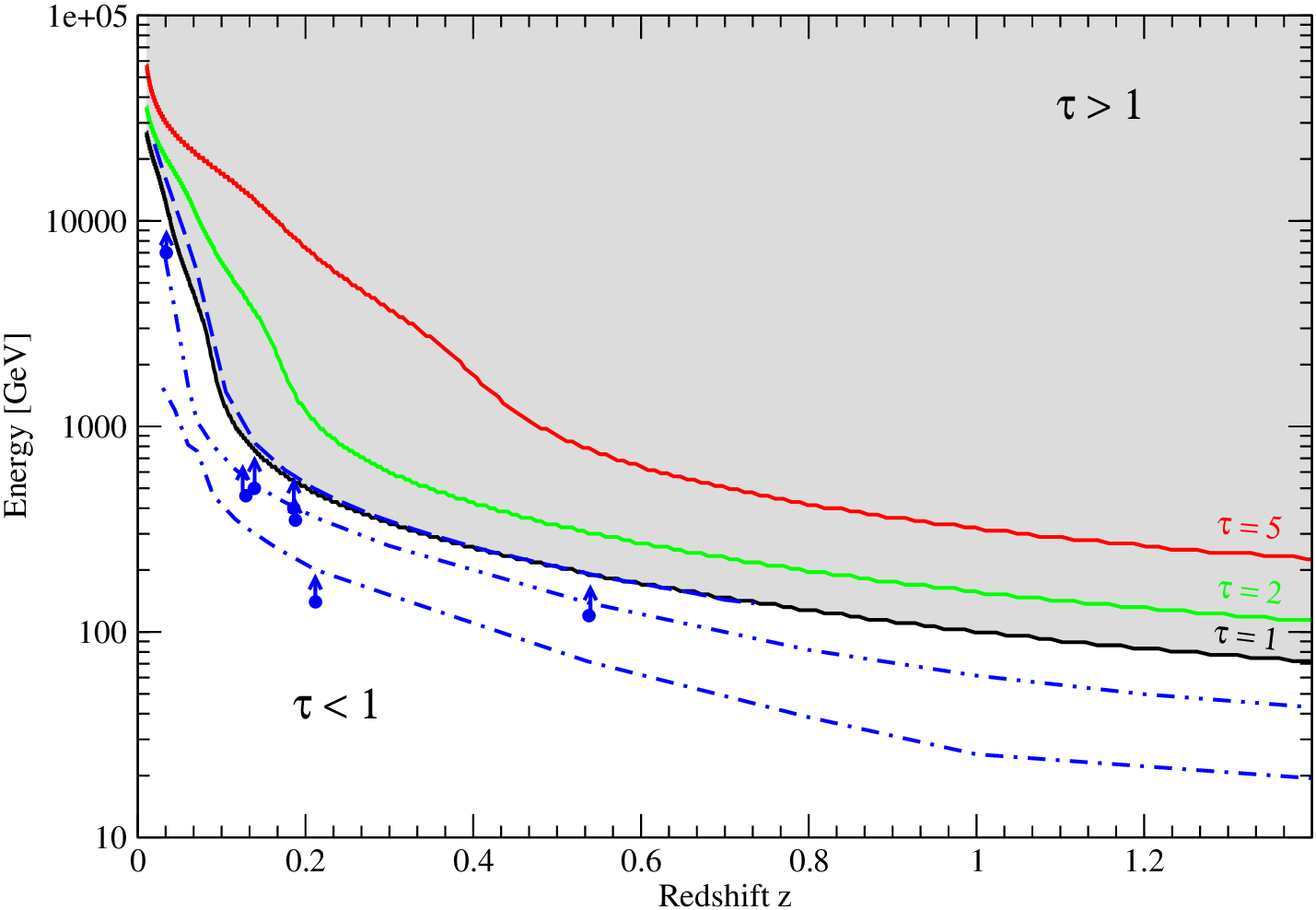} 
\caption{Gamma-ray horizon $\tau_{\gamma\gamma}(E_c, z) = 1 $ (black
  line), $\tau_{\gamma\gamma}(E_c, z) = 2 $ (green line),
  $\tau_{\gamma\gamma}(E_c, z) = 3 $ (red line) for the lower limit
  EBL model derived in this work.  Observed limits (dots) are
    taken from Fig.~3 of Albert et. (2008). For comparison, horizons
    based on three other EBL models are shown in blue, from the bottom:
    \cite{stecker2006}(dot-dashed), Albert et al. (2008)(dot-dot-dashed) and
    Primack (2005)(dashed).} 
\label{abb:horizon} 
\end{figure*}

The attenuation of gamma-rays can also be expressed by the
Fazio-Stecker relation, also known as the gamma-ray horizon. It is
shown in Figure~\ref{abb:horizon}, for a source-independent
description.  The redshift of a high energy gamma-ray source is
plotted against gamma-ray energy for an optical depth
$\tau_{\gamma\gamma}(E_c, z) = 1 $ (black line),
$\tau_{\gamma\gamma}(E_c, z) = 2 $ (green line),
$\tau_{\gamma\gamma}(E_c, z) = 3 $ (red line). This line are
calculated using the lower-limit model derived in this work.  Limits
from blazar observations are plotted as well taken from Albert et
al. (2008). The blazars all lie in the transparent region ($\tau <
1$), according to our model. For a given energy, blazars at a slightly
higher redshift than already measured might be detected.  All data are
in agreement with the lower limit model.  Despite the fact that a
lower-limit EBL has been used, there is little room left for a higher
EBL flux resulting in a higher optical depth for high energy
gamma-rays.

Finally the result is compared with models by Primack (2005), Albert
et al. (2008) and Stecker et~al. (2006) (dashed, dot-dot-dashed, and dot-dash
lines). Note that the EBL ''upper-limit'' model derived in Albert et
al. (2008) is based on the same code as presented here, but with a
completely different set of parameters, like star-formation rate, dust
and gas opacity etc. (see Table 1).  Our lower-limit model predicts
the smallest correction for extragalactic absorption, as expected,
except at very low redshifts ($z<0.2$), where the Primack (2005) model
is slightly above ours. This can be explained by the underestimation
in the far-infrared of this model, below the lower limits.


\section{Conclusions} 
In this paper a lower-limit EBL model has been derived utilising the lower limit
data from the integration of galaxy number counts from the optical to
the far infrared region. The model takes into account time-evolution
of galaxies, and includes the effect of absorption and re-emission of
the interstellar medium. To get such a low EBL, the assumption of a
quite low cosmic SFR has to be made, which has a maximum at a redshift
of 1.2 of about 0.1 M$_\odot$ yr$^{-1}$ Mpc$^{-3}$ and is falling to a
value of about 0.03 at a redshift of 5.
As expected the present-day lower-limit EBL is still
below the upper limits derived so far from the process of pair
production with very high energy gamma-ray emission by BL Lacs (see
red-dashed line in Fig.3).

This model can be used to calculate the interaction of cosmic-ray
particles with ambient photons fields. Cosmic-ray protons loose energy
due to pion production with stellar photons if their energy lies in
the range between $10^{16}$ and $10^{19}$~eV. Using the EBL model a
minimum, guaranteed energy loss of protons can be derived.

A lower-limit EBL model, is also essential to test exotic particle
physics scenarios in the universe. Particles, like Axions
(Sanchez-Conde et al. 2009) or hidden photons (Zechlin, Horns \&
Redondo 2008), can prevent high energy gamma-ray photons from being
absorbed. Other mechanisms like Lorentz invariance violations
(Protheroe \& Meyer 2000) can only be studied if the uncertainty of
the EBL is as small as possible.  A minimum absorption due to a
quaranteed low energy photon field from galaxies is essential to look
for such particles and effects.

In this paper it has been used to compute the absorption factor for
gamma-rays and observed blazar spectra at some selected redshifts. The
Fazio-Stecker relation which describes the absorption of high energy
gamma-rays from extragalactic sources as a function of redshift has
also been calculated. From this it can be concluded that the
lower-limit EBL flux can be used to correct high energy gamma-ray
spectra at all redshifts. The minimum correction done with this model
seems to lead to realistic intrinsic gamma-ray spectra of AGN even at
high redshift which can be modeled with standard acceleration
scenarios in relativistic jets. Up to now it was only possible to
show the agreement between lower-limit data and indirect upper limits
for the present day EBL flux. In this paper it was shown that also at
higher redshift only an EBL close to 
a lower-limit extragalactic diffuse photon flux,
taking into account the complete cosmic evolution of galaxies, is in
agreement with upper limits from high redshift blazar observations.

The recent detection of 3C279 blazar at $z=0.536$ by the MAGIC
collaboration (\cite{albert2008,errando2009}) has arisen the question
of the transparency of the Universe to the $\gamma$-rays, and the
level of the Cosmic Infrared Background (e.g. Aharonian et al. 2006,
Aharonian et al. 2007a, Stecker \& Scully 2009).  It can be confirmed
that the current lower limits of the EBL flux also at a redshift as
high as $z=0.536$ are fully compatible with $\gamma$-ray observations,
both on the blazar SED and on the $\gamma$-ray horizon.

If, in the future, EBL limits from TeV observations become lower,
maybe even dropping below the strict lower-limit EBL, the assumptions
leading to EBL limits from gamma-ray observations might have to be
revised.  On the other hand, the discovery of AGN showing a spectral
behavior which is not in agreement with our derived gamma-ray
horizon, would challenge AGN physics.

The lower limit EBL data, the EBL flux and optical depth as a function of
wavelength/energy and redshift are electronically
available\footnote{in Orsay: http://www.ias.u-psud.fr/irgalaxies/ and
  in Hamburg: http://www.astroparticle.de}.

\begin{acknowledgements} 
  We thank Andrew Hopkins for providing us with an electronic form of
  the CSFR compilation. We thank Wolfgang Rhode and Martin Raue for
  useful discussions.  TK acknowledges the support of DFG grant Kn
  765/1-2.  HD acknowledges the support of ANR-06-BLAN-0170.
\end{acknowledgements}

\appendix

\begin{figure*}
\includegraphics[width=0.75\textwidth]{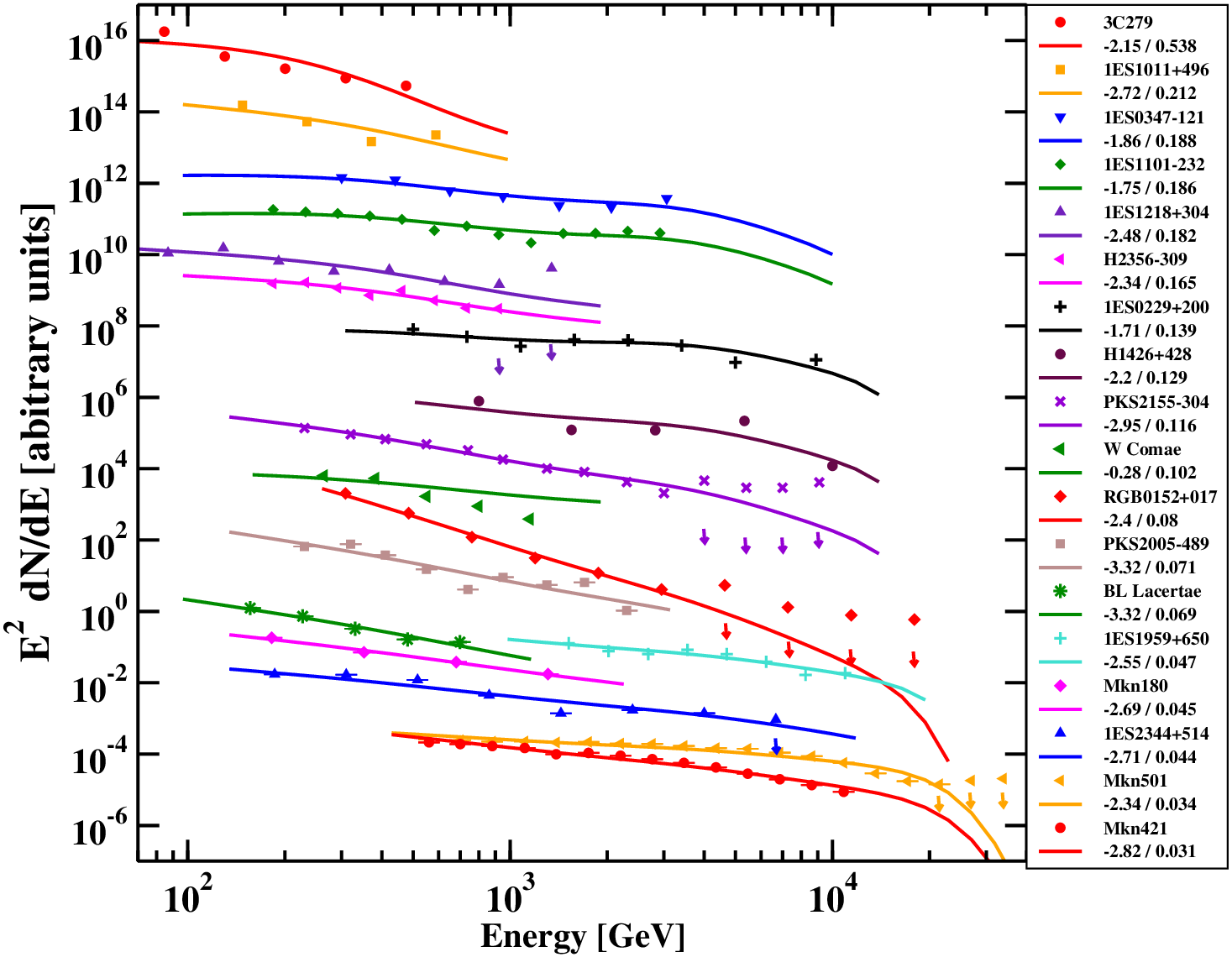} 
\caption{Observed spectral energy distributions for blazars (
    indicated at the right of the figure): dots (data), lines
  (model). The sources are ordered by their redshift, from high (top)
  to low redshift (bottom). The total flux is normalized for a better
  visualization. The lines are model spectra corrected for minimum EBL
  absorption, described in the text.  Numbers on the right indicate
  the spectral index $\alpha$ and the redshift of the source.  }
\label{abb:blazars} 
\end{figure*}  

\section{Application to the SED of Blazars}

The lower-limit EBL model is used to calculate spectral energy
distribution for observed TeV-blazars. To compare the spectra with the
observations, a single power-law is used with a spectral index
indicated below the source name in the table right to
  figure~\ref{abb:blazars}. Figure~\ref{abb:blazars} shows the
  spectra of blazars, sorted by increasing redshift (from bottom to
  top) and multiplied by an arbitrary constant to ease the
  visibility. 
The spectral index and normalization 
has been taken from a fit of the corrected data points of each source.
Then the powerlaw was multiplied by the extinction factor shown in Fig.~4 depending
on the redshift of the gamma-ray source. Using this method we get a continuous spectrum
for each source.

The intrinsic spectra can all be described by power-laws with
spectral indices still in agreement with very simple jet models in AGN, like the
synchrotron-self compton model (SSC). 
This was not surprising, given the
  lower limit EBL which has been used. But this might be another indication that the
  opacity to $\gamma$-rays is still low ($\tau < 1$), even at higher redshift
  $z \sim 0.5$.


\end{document}